\documentclass[a4paper,11pt]{article}
\usepackage{pos}
\usepackage{url}

\title{The Cherenkov Telescope Array transient and multi-messenger program}
 \ShortTitle{CTA multi-messenger program}

\author*[a]{Alessandro Carosi}
\author[b]{Alicia L\'opez-Oramas}
\author[c]{Francesco Longo}

\affiliation[a]{DPNC – University of Geneva 24 quai Ernest Ansermet, CH1211 Genève 4, Switzerland}
\affiliation[b]{Instituto de Astrofisica de Canarias and Universidad de La Laguna,
 La Laguna, Spain}
\affiliation[c]{University of Trieste and INFN, Trieste, Italy}

\forColl{CTA} 

\emailAdd{alessandro.carosi@unige.ch}

\abstract{The Cherenkov Telescope Array (CTA) is a next generation ground-based very-high-energy gamma-ray observatory that will allow for observations in the >10 GeV range with unprecedented photon statistics and sensitivity. This will enable the investigation of the yet-marginally explored physics of short-time-scale transient events. CTA will thus become an invaluable instrument for the study of the physics of the most extreme and violent objects and their interactions with the surrounding environment. The CTA Transient program includes follow-up observations of a wide range of multi-wavelength and multi-messenger alerts, ranging from compact galactic binary systems to extragalactic events such as gamma-ray bursts (GRBs), core-collapse supernovae and bright AGN flares. In recent years, the first firm detection of GRBs by current Cherenkov telescope collaborations, the proven connection between gravitational waves and short GRBs, as well as the possible neutrino-blazar association with TXS~0506+056 have shown the importance of coordinated follow-up observations triggered by these different cosmic signals in the framework of the birth of multi-messenger astrophysics. In the next years, CTA will play a major role in these types of observations by taking advantage of its fast slewing (especially for the CTA Large Size Telescopes), large effective area and good sensitivity, opening new opportunities for time-domain astrophysics in an energy range not affected by selective absorption processes typical of other wavelengths. In this contribution we highlight the common approach adopted by the CTA Transients physics working group to perform the study of transient sources in the very-high-energy regime.}

\FullConference{37$^{\rm{th}}$ International Cosmic Ray Conference (ICRC 2021)\\
		July 12th -- 23rd, 2021\\
		Online -- Berlin, Germany}

\begin{document}
\maketitle

\section{Introduction}
\label{sec:intro}
In recent years, the field of very-high-energy (VHE, E$>$100 GeV) transient astronomy has begun evolving evolving toward the observations of new sources in the context of a multi-messenger approach. The comprehension of the connection between time-domain and multi-messenger astrophysics is rapidly becoming of primary importance since VHE photons are produced in extreme, violent environments potentially associated with cosmic accelerators and therefore to the production of high-energy cosmic-rays and neutrinos. Furthermore, VHE radiation might be expected from stellar collapses and compact object mergers which power supernovae, gamma-ray bursts (GRBs) and gravitational wave (GW) emission. Transients are an integral part of the CTA {\it Key Science Projects} (KSP)~\citep{sciencewithcta}. A dedicated Science Working Group (Transient and multi wavelength SWG) is in place to prepare first observations (react to fast target of opportunities-ToO, define the observation program, prepare the scientific analysis, etc..) and set up the needed multi wavelength/multi-messenger connections and synergies with external facilities. The main scientific goals of the group, include the release of dedicated consortium publications focused on key topics such as GRBs, gravitational waves, neutrino ToOs, galactic transients and core-collapse supernovae. The group is also involved in other activities such as evaluating the detection prospects of serendipitous VHE transients identified via the CTA real-time analysis and the VHE transient survey, by exploring the divergent pointing capability in association with the CTA extragalactic survey KSP.

In this contribution, we will report on the general Transients KSP and each of the main sub-projects (namely consortium publications) under development within the Transients SWG, their current status and planned activities. 

\section{The Transients KSP}
\label{sec:ksp}
\begin{figure}
\centering
\includegraphics[width=0.42\columnwidth]{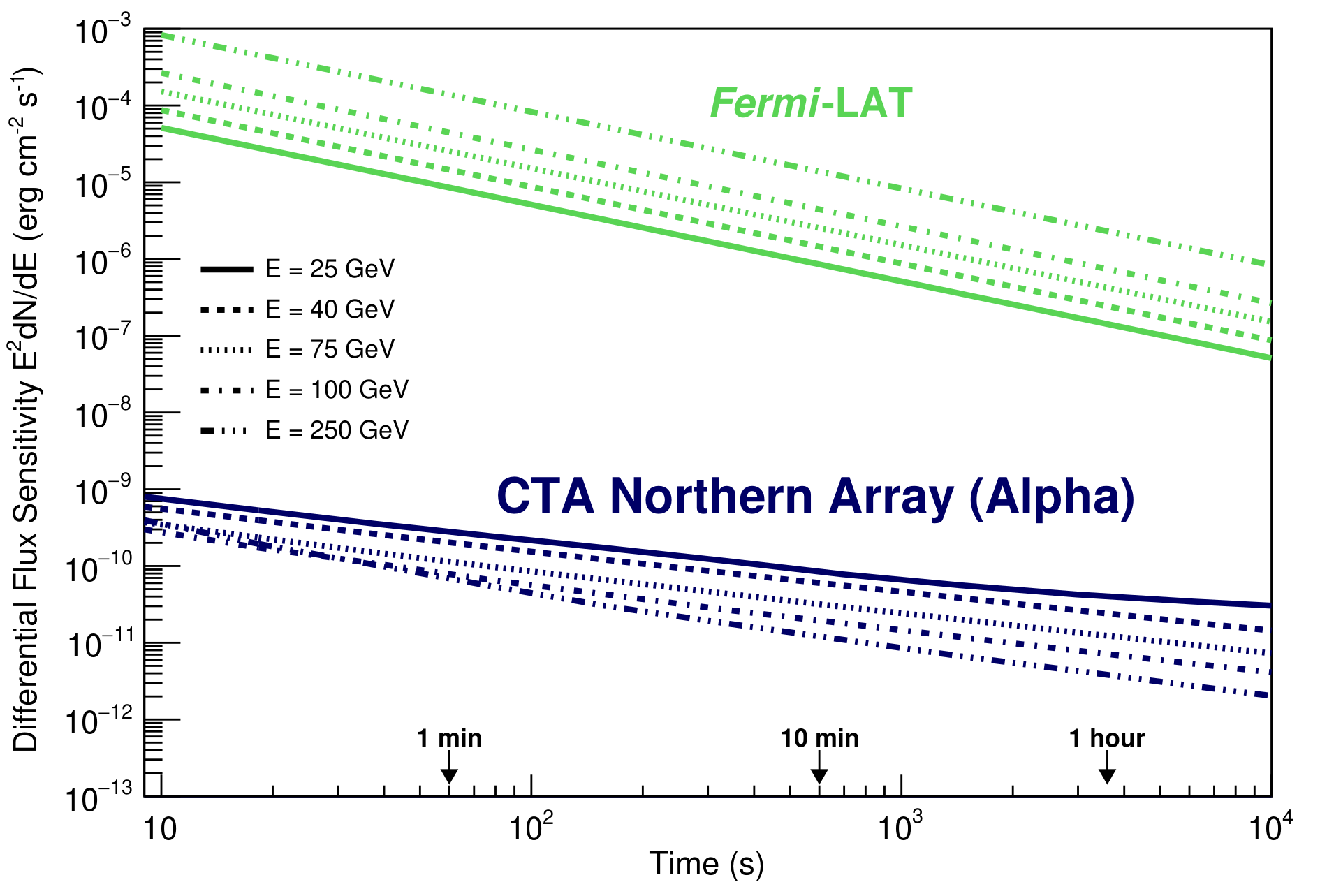}
\includegraphics[width=0.55\columnwidth]{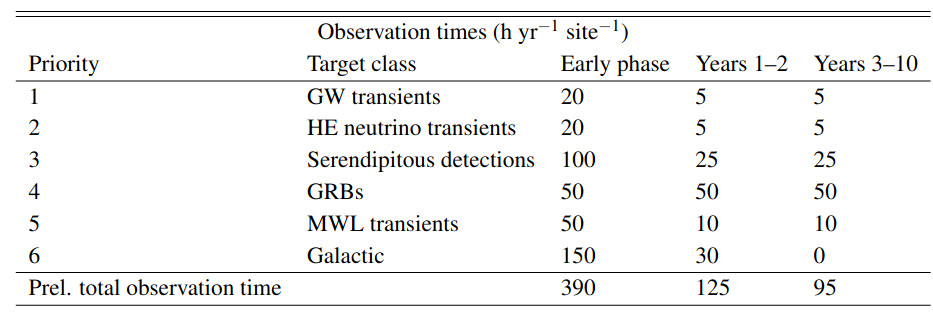}
\caption{{\it Left:} Sensitivities of CTA and {\it Fermi}-LAT in their overlapping energy range as a function of observation time. Differential flux sensitivities at three energies are compared. For CTA, the proposed \emph{alpha configuration} with 4 large sized telescopes (LST) and 9 medium sized telescopes (MST) in the north array is reported. From~\url{https://www.cta-observatory.org/science/cta-performance/}. {\it Right:} Summary of the observation proposed in the Key Science Project proposal in~\citep{sciencewithcta}.}
\label{fig:stsens}
\end{figure}

The Transients KSP is an important part of the CTA science core program and one of its main goals is to prepare the response of CTA to a wide range of multi-wavelength and multi-messenger alerts, including GRBs, GWs and high-energy neutrinos. The sensitivity achievable to emission on short timescales and the fast response to external alerts will allow CTA to probe ultra-fast variability over early phases of transient events in unprecedented detail compared to current generation imaging atmospheric Cherenkov telescopes (IACTs) and space-based instrumentation (see Fig.\ref{fig:stsens} left panel). 

This would allow us to shed some light on some fundamental topics such as:

\begin{itemize}
\item The investigation of the physical mechanisms that drive jets and winds around neutron stars and black holes
\item The study of and potential discrimination between the different emission scenarios and physical processes capable of producing the observed HE and VHE radiation from GRBs
\item The origin of the ultra-high energy cosmic rays (UHECRs) and high-energy neutrinos
\item The study of the physical mechanisms at the origin of GWs and their electromagnetic counterparts
\end{itemize}

Starting from these key science topics, and also using the extensive experience of current IACTs, the transient KSP is poised to optimize the CTA response to transient alerts by providing a list of detailed use cases covering a wide range of observational steps: from the receiving of the alerts to the reaction of the arrays; from data taking to the rapid feedback to the astronomical community on the VHE properties of the observed transients. A preliminary list of the proposed observations within the CTA transient KSP is reported in Fig.(\ref{fig:stsens}) although the outlined numbers are currently going under a major revision according to the latest, state-of-the-art knowledge derived from current IACTs follow-ups.

\section{Gamma-ray Bursts}
\label{sec:grb}
GRBs have always been considered a primary targets for all modern IACT telescopes. The detection of VHE gamma-ray from a number of events including GRB~180720B~\cite{HESS_GRB180720B}, GRB~190114C~\cite{MAGIC_GRB190114C_1} and GRB~190829A~\cite{HESS_GRB190829A} represents a long-awaited result and a remarkable step forward in our understanding of GRB physics. For long time, the search for VHE signals associated with GRBs posed a major challenge for IACTs from both the technical and the scientific point of view (see e.g.~\cite{Covino10}~\cite{Carosi13}). The possibility of detecting a VHE gamma-ray signal from a GRB is indeed crucial for clarifying the poorly-known physics of these objects during the different phases of their emission, in particular during the early afterglow phase when the co-existence of forward and reverse shocks in the emitted outflow could yield a large variety of different emitting scenarios. The CTA array will routinely perform follow-up observations of GRB triggers. The estimation of the detection prospects for such observations are necessarily still preliminary and are dependent on the final array layout and performance. However, even starting with simplified assumptions about the GRB emission, the CTA Consortium already reported the possibility of detecting $\sim$hundreds (or more) of photons from moderate to bright GRB allowing for a significant improvement in the photon statistics and for the possibility to have good-quality time-resolved spectra~\citep{Inoue2013}. In order to achieve a step forward in the determination of CTA's prospects for GRB follow-ups, the Transient SWG is currently working on a new publication where the potential detection rate is estimated using a theoretical-based approach. Such an approach is based on the \emph{POpulation Synthesis Theory Integrated code for Very high energy Emission} (POSyTIVE) population model for GRBs~\citep{posytive}. The resulting population is built by considering few intrinsic properties and assumptions:

\begin{itemize}
    \item E$_{\rm peak}$ \& redshift distribution
    \item E$_{\rm peak}$-E$_{\rm iso}$ correlation (Amati relation)~\cite{amati}
    \item Bulk Lorentz factor distribution obtained by measured the time of the afterglow onset (providing the bulk Lorentz factor of the event's coasting phase)
\end{itemize}

The population obtained (for both long and short GRBs) is calibrated against a wide data set of multi-wavelength observations. In order to derive the final expected spectrum,  both the prompt and the afterglow emission are simulated according to standard leptonic synchrotron and synchrotron self-Compton models~\cite{grb_emission}. The GRB spectra obtained are then used to simulate the detailed CTA response through the use of dedicated analysis pipelines based on gammapy\footnote{\url{https://gammapy.org/}} and ctools\footnote{\url{http://cta.irap.omp.eu/ctools/}}.    

\begin{figure}
\centering
\includegraphics[width=0.59\columnwidth]{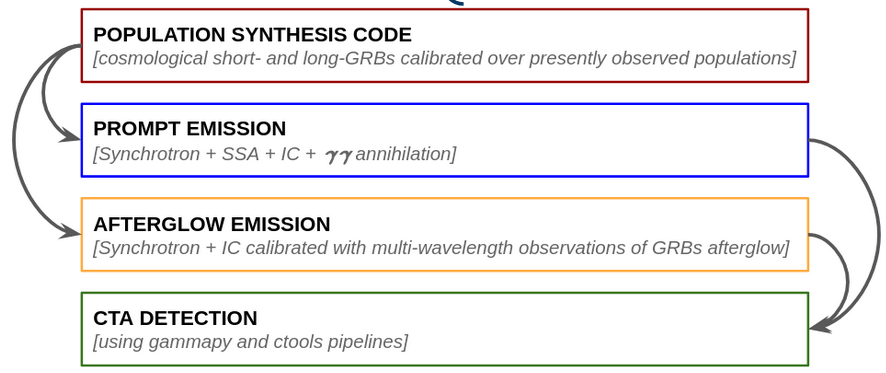}
\includegraphics[width=0.4\columnwidth]{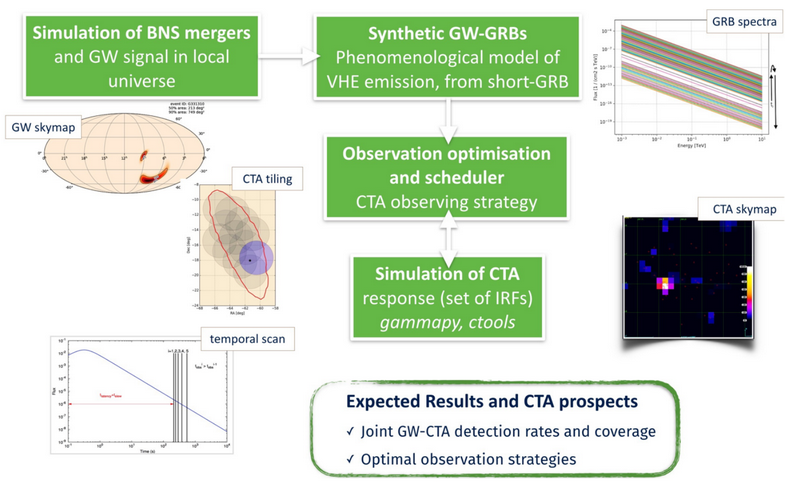}
\caption{({\it Left:}) Scheme of the GRB consortium publication work. Synthetic spectra and light curves are obtained by a population synthesis code and used to feed the CTA analysis pipeline. ({\it Right:}) Workflow for the GW consortium publication. After the simulation of BNS merger, a phenomenological (short) GRB is associated to it. The optimal pointing strategy is then obtained by a dedicated algorithm in order to cover efficiently the sky area of the GW source. Each pointing is then analyzed  by means of the CTA analysis pipeline.}
\label{fig:grb_scheme}
\end{figure}

\section{Gravitational Waves}
\label{sec:gw}
The link between GWs and short GRBs (sGRBs) has been discussed widely in literature in the past and was proven definitively after the observation of the coincidence between GRB\,170817A and the gravitational signal GW\,170817~\citep{GW-sGRB}. Although all of the GRBs detected so far by current IACTs were long GRB, sGRBs are also expected to emit VHE radiation. Nevertheless, only a hint of emission in this energy band are reported so far for sGRBs by the MAGIC telescopes~\citep{magic_sgrb}. However, this result, as well as the significant programs put in place by the major IACT collaborations~\citep[see e.g.,][]{hess_gw1,hess_gw2} for follow-up campaigns of GW triggers confirms the scientific potential of these observations. Thanks  to  its  unprecedented sensitivity  and  the  increased  number  of  GW events expected in the near future, CTA will be able to increase the number of VHE counterpart detections providing a deeper insight into their physical processes. The detection of the potential electromagnetic counterpart is, however, challenging due to the relatively large localisation uncertainties provided by GW interferometers. Within the transient SWG, a detailed study to establish the number of possible successful CTA detections of VHE counterparts of GWs is under development~\citep{cta_gw}. 

In contrast to the GRB case, a purely phenomenological approach is used: a short GRB is associated to a set of simulated binary neutron star (BNS) mergers extracted from the public database GWCOSMos~\citep{gwcosmos} providing the GW skymap, distance and orientation. The corresponding VHE emission is derived from the empirical correlations between the X-ray and TeV luminosities as observed in the VHE GRB sample (as in GRB\,190114C). The optimal pointing strategy is then obtained by a dedicated algorithm in order to cover efficiently the sky area of the GW source. Each pointing is subsequently analyzed by means of a dedicated analysis pipeline in order to provide the final outcome on a possible detection. The CTA GW follow-up program is currently being defined and implemented. The results of these studies will be the subject of a dedicated consortium publication expected by the end of 2021.

\section{Neutrino Follow-up}
\label{sec:neutrino}

\begin{figure}
\centering
\includegraphics[width=0.45\columnwidth]{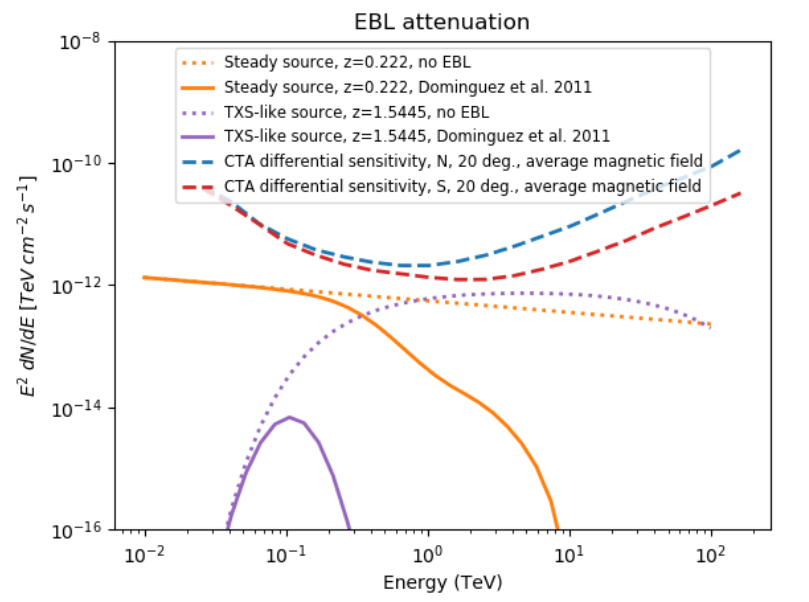}
\includegraphics[width=0.45\columnwidth]{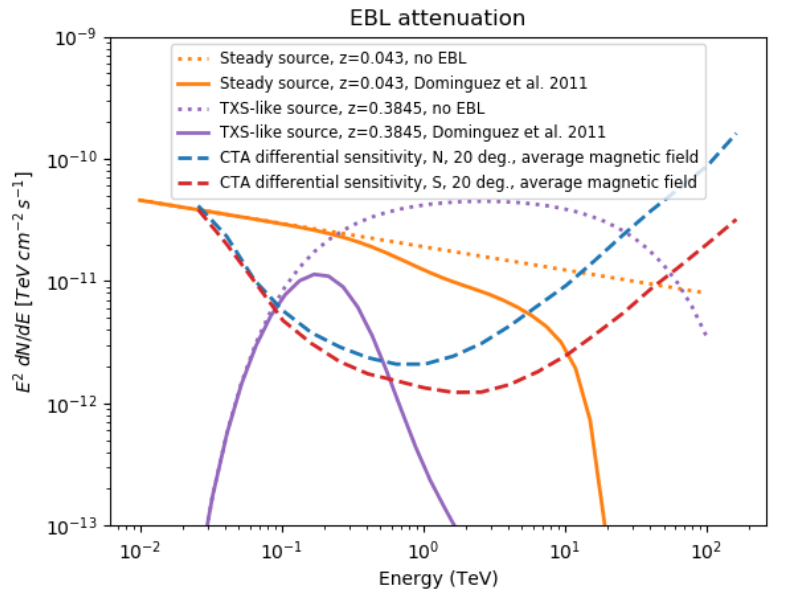}
\caption{The energy spectra of potential neutrino sources (without and with the EBL attenuation) overlaid on the CTA differential sensitivity for the undetected (left) and detected (right) sources. CTA sensitivities correspond to the \emph{omega configuration} for both the Southern (4 LSTs + 25 MSTs + 70 SSTs) and the Northern hemisphere (4 LSTs + 15 MSTs) from~\citep{whitepaper}.}
\label{fig:neutrino_detection}
\end{figure}

The IceCube detection of the high-energy neutrino event (IC-170922A) associated with the flaring gamma-ray blazar TXS 0506+056 represents the best evidence so far seen for an astrophysical neutrino point source~\citep{IC170922A}. The extensive multi-wavelength follow-up campaign that also involved many of the currently operating  IACTs, stands as an important test case for the optimization of the CTA Neutrino Target of Opportunity (NToO) program. The aim of this program is to develop a strategy for CTA follow-up of neutrino alerts to maximize the chance of detecting a VHE counterpart. Neutrino point source simulations are based on FIRESONG~\citep{firesong}, which takes into account the cosmological evolution of different source classes and the recent results from IceCube (i.e., the measured diffuse flux of astrophysical neutrinos). These are then the input for simulating the expected VHE gamma-ray emission. Neutrinos might indeed be accompanied by VHE gamma rays produced according to typical pp and p$\gamma$ models. CTA, with its fast reaction time and lower energy threshold, will enable sensitive searches for such VHE counterparts for well-localized, likely astrophysical neutrino events up to much higher redshifts than those accessible to current IACTs (Fig.~\ref{fig:neutrino_detection}).  Preliminary results~\citep{nucta1, whitepaper} show that, for flaring blazars, CTA will be able to detect a VHE counterpart for about one third of the cases after $\sim$10 mins of observations, with lower detection probabilities for steady neutrino sources. For more details, see the dedicated contribution at this conference~\citep{nucta2}.

\section{Galactic Transients}
\label{sec:galactic}

Several sources in our Galaxy exhibit transient emission via different processes, such as outflows interacting with the surrounding interstellar medium, strong winds or accretion/ejection from/onto a compact object. Some of these events are energetic enough to accelerate particles via non-thermal mechanisms and produce high-energy gamma-ray emission in the MeV range. But the main question that arises is whether these MeV emitters can also be sources of VHE gamma rays. For the moment, searches for VHE emission from different types of gamma-ray emitters have not revealed any signal with current IACTs. 
CTA, with its unprecedented sensitivity to short-timescale transient events, will perform first-time detections of the VHE component of many MeV-emitters and will likely achieve serendipitous discoveries of yet unknown sources. CTA observations of these flaring objects will be triggered by external facilities such as X-ray or high-energy monitoring satellites. Serendipitous detections might also happen while performing, i.e., surveys such as that of the Galactic Plane. For the detection of Galactic transients, the real-time analysis will play a key role in the follow-up and observation strategies of these externally triggered events. Among the different types of Galactic transients, we have tested the capabilities of CTA to detect transient emission from two types of MeV-emitters: flares from pulsar wind nebulae (PWNe) and microquasars. A more detailed analysis of these and others Galactic transient sources, is reported in~\citep{aliciaICRC21}.

\textbf{Flares from the Crab Nebula: }
PWNe are bubbles of relativistic plasma which are powered by the magnetically-driven winds of a central highly-magnetized rotating neutron star (pulsar). The Crab Nebula PWN is the brightest persistent VHE source in the TeV gamma-ray sky, and is referred to as the standard candle. AGILE, however, discovered flaring activity from this source~\cite{Tavani2011Sci...331..736T}, which was subsequently confirmed by {\it Fermi}-LAT \citep{Abdo2011}. We prove that CTA-N will be able to detect VHE flaring emission from the Crab Nebula down to $10\%$ of flare intensity measured by AGILE, thanks to its wide energy range and its improved sensitivity with respect to the current generation of IACTs. 

\textbf{Microquasars: }
Microquasars are binaries composed of a compact object (which can be a black hole or a neutron star) and a companion star. Matter is accreted onto the compact object, creating an accretion disk and (eventually) generating collimated jets of plasma. We have tested whether some MeV emitters, such as the microquasars in the Cygnus region, can also be detected at VHE by CTA. We have tested whether the two massive microquasars  Cyg X-1 and Cyg X-3, which are known MeV sources \cite{AGILE2010ApJ...712L..10S, Zanin2016} could also be detected with CTA-N, even if no evidence for VHE emission has been claimed by current IACTs \cite{Aleksic2010,Ahnen2017}. Only marginal evidence for a VHE signal was observed by the MAGIC telescopes during an 80-minute long observation in 2006 \cite{Albert2007}, coincident with a hard X-ray flare. We highlight the capability of the CTA-N array to detect transient emission from both of the microquasars, emphasizing the detection of transient emission from Cyg X-1 in only 30 minutes. 

CTA will  provide crucial insights which will help to refine emission models for microquasars and jet formation, and will reveal the mechanisms at work in flaring PWNe at VHE. Its unique sensitivity to short-time events and its large energy coverage will allow CTA to discover the VHE components of many Galactic MeV emitters and to reveal serendipitous detections of unknown sources.

\section{Core-Collapse Supernovae}
\label{sec:ccsn}
A core-collapse supernova (CCSNe) represents the catastrophic explosion of a massive star at the end of its life. The energy is mainly released in the form of kinetic energy of a  non-relativistic expanding outflow. In the resulting fast-moving shock wave, particles are accelerated via the first-order Fermi mechanism. The accelerated particles, interacting with the surrounding interstellar matter, might lead to the production of a gamma-ray signal up to the VHE band that can be potentially detected by CTA. A wide range of different types of CCSNe exists (IIP, IIL, IIb, IIn, etc. ) which could each have a different signature in the VHE regime. A dedicated work on the  prospects for the CTA detection of such a signal has recently started within the Transients working group. A set of known CCSne will be used as a template to model the expected VHE emission in the CTA energy range. Such emission is then input to dedicated CTA analysis pipelines to study the CTA prospects for such observations during the first days/months (and possibly years) after the explosion.

\section{Acknowledgements}
This work was conducted in the context of the CTA Transients Working Group. We gratefully acknowledge financial support from the agencies and organizations listed here: \url{http://www.cta-observatory.org/consortium\_acknowledgments}.

\section*{Full Authors List: \Coll\ Collaboration}
H.~Abdalla\textsuperscript{1}, H.~Abe\textsuperscript{2},
S.~Abe\textsuperscript{2}, A.~Abusleme\textsuperscript{3},
F.~Acero\textsuperscript{4}, A.~Acharyya\textsuperscript{5}, V.~Acín
Portella\textsuperscript{6}, K.~Ackley\textsuperscript{7},
R.~Adam\textsuperscript{8}, C.~Adams\textsuperscript{9}
% ...

1 : Centre for Space Research, North-West University, Potchefstroom, 2520, South Africa

2 : Institute for Cosmic Ray Research, University of Tokyo, 5-1-5, Kashiwa-no-ha, Kashiwa, Chiba 277-8582, Japan

3 : Pontificia Universidad Católica de Chile, Av. Libertador Bernardo O'Higgins 340, Santiago, Chile

4 : AIM, CEA, CNRS, Université Paris-Saclay, Université Paris Diderot, Sorbonne Paris Cité, CEA Paris-Saclay, IRFU/DAp, Bat 709, Orme des Merisiers, 91191 Gif-sur-Yvette, France

5 : Centre for Advanced Instrumentation, Dept. of Physics, Durham University, South Road, Durham DH1 3LE, United Kingdom

6 : Port d'Informació Científica, Edifici D, Carrer de l'Albareda, 08193 Bellaterrra (Cerdanyola del Vallès), Spain

7 : School of Physics and Astronomy, Monash University, Melbourne, Victoria 3800, Australia

8 : Laboratoire Leprince-Ringuet, École Polytechnique (UMR 7638, CNRS/IN2P3, Institut Polytechnique de Paris), 91128 Palaiseau, France

9 : Department of Physics, Columbia University, 538 West 120th Street, New York, NY 10027, USA

% ...

%
%
%


\begin{thebibliography}{99}
\bibitem{sciencewithcta} B.~S.~Archaya et al (CTA Consortium), 2019, Published by World Scientific Publishing Co., https://doi.org/10.1142/10986
\bibitem{HESS_GRB180720B} H.E.S.S. Collaboration: Abdalla, H. \& et al., 2019, Nature, 575, 464
\bibitem{MAGIC_GRB190114C_1} MAGIC Collaboration: Acciari, V. A. \& et al., 2019, Nature, 575, 455
\bibitem{HESS_GRB190829A} H.E.S.S. Collaboration: ATeL n.13052
\bibitem{Covino10} MAGIC Collaboration, Aleksi\'c, J. et al., 2010, A\&A 517, 5
\bibitem{Carosi13} MAGIC Collaboration, Aleksi\'c, J. et al., 2013, MNRAS, 437, 3103
\bibitem{Inoue2013} Inoue, S., Granot, J., O'Brien, P. et al., 2013,  APh, 43, 252
\bibitem{posytive} Sadeh,~I. et al., 2019, in 36th International Cosmic Ray Conference (ICRC~2019), PoS598
\bibitem{amati} Amati, ~L., et  al., 2002, A\&A, 390, 81 
\bibitem{grb_emission} R.~Sari \& A.~A.~Esin, 2001, ApJ, 548, 787
\bibitem{GW-sGRB} B. P. Abbott,~B.~P. R. Abbott,~R., Abbott,~T.D. et al., 2017, \emph{ApJL}, {\bf 848} L12
\bibitem{magic_sgrb} MAGIC Collaboration: Acciari, V.~A. et al, 2021, Astrophys. J.908, 90, 2012.07193.
\bibitem{hess_gw1} H.E.S.S. Collaboration: Abdalla,~H. et al., 2017, Astrophys. J. Lett.850, L22, 1710.05862
\bibitem{hess_gw2} H.E.S.S. Collaboration: Abdalla,~H. et al., 2020, Astrophys. J. Lett.894, L16, 2004.10105.
\bibitem{cta_gw} Seglar-Arroyo,~M. et al., 2019, in 36th International Cosmic Ray Conference (ICRC 2019), PoS790.
\bibitem{gwcosmos} Patricelli,~B., Stamerra,~A., Razzano,~M. et al. 2018, JCAP, 05, 056
\bibitem{IC170922A} IceCube Collaboration, 2018, Science 361, 1378
\bibitem{firesong} Taboada,~I. et al., 2019, in 35th International Cosmic Ray Conference (ICRC 2017) PoS(ICRC2017)663
\bibitem{nucta1} Satalecka,~K., Brown,~A., Rosales de Leon, O. Sergijenko, et al. (CTA Consortium, FIRESONG Team), in 36th International Cosmic Ray Conference (ICRC 2019)(2019), vol. 36 of International Cosmic Ray Conference, p. 784.
\bibitem{whitepaper} Bosnjak,~Z., Brown,~A.M., Carosi,~A. at al. (CTA Consortium), 2021, Submitted to ASTRONET roadmap, arXiv:2106.03621v
\bibitem{nucta2} O.~Sergijenko et al. (CTA Consortium), 2021, in 37th International Cosmic Ray Conference (ICRC 2021), contribution id.224
\bibitem{aliciaICRC21} L\'opez-Oramas,~A. et al. ICRC2021,  (CTA Consortium), 2021, in 37th International Cosmic Ray Conference (ICRC 2021), contribution id.224
\bibitem{Tavani2011Sci...331..736T} Tavani,~M., Bulgarelli,~A., Vittorini,~V. et al., 2011, Science, Volume 331, Issue 6018
\bibitem{Abdo2011} Abdo~.A.~A. et al. (Fermi-LAT), 2011, Science 331, 739, 1011.3855
\bibitem{AGILE2010ApJ...712L..10S}  Sabatini,~S., Tavani,~M., Striani,~E., 2010, ApJL, 712, Issue 1
\bibitem{Zanin2016} Zanin,~R. et al. 2016, A\&A, 596, id.A55
\bibitem{Aleksic2010} MAGIC Collaboration: {Aleksi{\'c}},~J. et al., 2010, ApJ, 721, 1 
\bibitem{Ahnen2017} MAGIC Collaboration: Ahnen,~A. et al., 2017, MNRAS, 472, 3
\bibitem{Albert2007} MAGIC Collaboration: Albert,~A. et al., 2007, ApJL, 665, L51


\end{thebibliography}
\end{document}